\documentclass[10pt,letterpaper]{article}
\usepackage[top=0.85in,left=2.75in,footskip=0.75in]{geometry}

\usepackage{amsmath,amssymb}

\usepackage{changepage}

\usepackage[utf8x]{inputenc}

\usepackage{textcomp,marvosym}

\usepackage{cite}

\usepackage{nameref,hyperref}

\usepackage[right]{lineno}

\usepackage{microtype}
\DisableLigatures[f]{encoding = *, family = * }

\usepackage[table]{xcolor}

\usepackage{array}

\newcolumntype{+}{!{\vrule width 2pt}}

\newlength\savedwidth

\raggedright
\usepackage[aboveskip=1pt,labelfont=bf,labelsep=period,justification=raggedright,singlelinecheck=off]{caption}

\makeatletter
\renewcommand{\@biblabel}[1]{\quad#1.}
\makeatother

\usepackage{lastpage,fancyhdr,graphicx}
\usepackage{epstopdf}
\fancyheadoffset[L]{2.25in}
\fancyfootoffset[L]{2.25in}
\begin{document}
\vspace*{0.2in}

\begin{flushleft}
{\Large
\textbf\newline{Non-Markovian intracellular transport with sub-diffusion and run-length dependent detachment rate} 
}
\newline
\\
Nickolay Korabel\textsuperscript{1*},
Thomas A. Waigh\textsuperscript{2,3},
Sergei Fedotov\textsuperscript{1},
Viki J. Allan\textsuperscript{4}
\\
\bigskip
\textbf{1} School of Mathematics, University of Manchester, Manchester, UK
\\
\textbf{2} Biological Physics, School of Physics and Astronomy, University of Manchester, Manchester, UK
\\
\textbf{3} The Photon Science Institute, University of Manchester, Manchester, UK
\\
\textbf{4} Faculty of Biology, Medicine and Health, School of Biological Sciences, University of Manchester, Manchester, UK
\\
\bigskip

%
%





* Email: nickolay.korabel@manchester.ac.uk

\end{flushleft}
\section*{Abstract}
Intracellular transport of organelles is fundamental 
to cell function and health. The mounting evidence suggests that this transport is in fact anomalous. However, the reasons for the anomaly is still under debate. We examined experimental trajectories of organelles inside a living cell and propose a mathematical model that describes the previously reported transition from sub-diffusive to super-diffusive motion. In order to explain super-diffusive behaviour at long times, we introduce non-Markovian detachment kinetics of the cargo: the rate of detachment is inversely proportional to the time since the last attachment. Recently, we observed the non-Markovian detachment rate experimentally in eukaryotic cells. Here we further discuss different scenarios of how this effective non-Markovian detachment rate could arise. The non-Markovian model is successful in simultaneously describing the time averaged variance (the time averaged mean squared displacement corrected for directed motion), the mean first passage time of trajectories and the multiple peaks observed in the distributions of cargo velocities. We argue that non-Markovian kinetics could be biologically beneficial compared to the Markovian kinetics commonly used for modelling, by increasing the average distance the cargoes travel when a microtubule is blocked by other filaments. In turn, sub-diffusion allows cargoes to reach neighbouring filaments with higher probability, which promotes active motion along the microtubules.


\section*{Introduction}

A variety of large structures and assemblies inside living cells, such as organelles, are now thought to experience sub-diffusive and/or super-diffusive motion rather than diffusive, Brownian motion \cite{MK,Granick,HF,Sokolov,Denisov,Barkai,Bressloff,Bressloff1,Rogers,Tabei,Jeon,Tolic,Golding,Caspi,Wang13,Weber,Reverey}. Thus the successful statistical models that describe Brownian motion developed by Einstein and Smoluchowski (among others) to describe 
the thermal fluctuations in the position of dilute colloidal particles (first demonstrated by J.B. Perrin) need to be adapted to describe motion in congested environments. Furthermore, this congested thermal motion is superposed on the non-equilibrium active transport that drives long distance organelle movement inside the cell \cite{Hirokawa}. This organelle transport is essential for cell function. Active organelle transport is
mediated by molecular motors which move processively along microtubules in the direction of their polarity. As an example, cytoplasmic dynein moves towards the minus ends of microtubules \cite{Allan, Allan1}. Molecular motors move by hydrolysing ATP molecules at rates of the order $100$ $s^{-1}$. The functioning of molecular motors may involve multiple states and it is regulated by a complex biochemical network \cite{Julicher,Kolomeisky}. 
It is believed that multiple motors moving collectively are required to transport large organelles 
\cite{KKBL,Lipowsky,Vershinin,KlumppLipowsky,Berger11,Berger12}.
However, the actual number of engaged motors {\it in vivo} is still unknown, although some fluorescence microscopy experiments indicate there can be up to $5$ dyneins and $5$ kinesins engaged simultaneously on large cargoes \cite{Muller}, and these numbers are expected to be dependent on the cargo type, cell type and the organism considered. Crucial unsolved questions in cellular biology include how anomalous motion impacts on the majority of processes 
inside a cell (both during mechanical interactions and in chemical reactions), how these processes can be modelled 
and how sub-diffusive transport combines with active motility driven by motor proteins \cite{Waigh2017}.

Recently, we experimentally found that the dynamics of cargo-motors complexes in eukaryotic cells is better described by a model with an effective non-Markovian kinetics \cite{SUBMITTED}. Namely, we found that the rate of detachment is inversely proportional to the time since the last attachment. Consequently the longer a cargo moves along a microtubule, the less likely it will detach from it. As a result, the movement of cargoes is non-Markovian and involves a memory. In this paper we analyze individual lipid droplet tracks \cite{Kenwright} and compare with the non-Markovian and the Markovian models of cargo transport. We focus on implications of non-Markovian dynamics for the physical properties of cargo transport. 
We show that in contrast to the Markovian model, the non-Markovian model is able to describe the process of super-diffusion previously reported in Ref. \cite{Kenwright}. 

The motion of the organelles in Ref. \cite{Kenwright} was observed to be often paused, displaying a run-rest behaviour. The nature of the rest states is not fully understood. Pauses can 
be caused by detachments of the motor from the microtubule. However, organelles were not found to diffuse away 
from the microtubule. Cytoplasmic dynein binds to the microtubule via a repeated cycle of detachment and reattachment. 
It was previously found that association with accessory proteins like dynactin may promote additional weak binding to the 
microtubule \cite{King,Kardon}. This prevents the motor-cargo complex from diffusing away while the dynein motor is 
detached, and facilitates the reattachment. Theoretically it was suggested that switching between free diffusion and 
active transport could enhance reaction kinetics \cite{Klumpp,Loverdo} and significantly improve signalling precision \cite{Godec1,Godec2}.

The behaviour of the time averaged mean squared displacement (TAMSD) of organelles was  found to be sub-diffusive 
at short time intervals $\Delta\ll1$ s in an experiment with liquid droplets \cite{Kenwright}, TAMSD $\sim \Delta^{\alpha}$ 
with $\alpha<1$. The exact physical origin of sub-diffusion is still under investigation \cite{Saxton,Saxton1,Scholz,Metzler14,Metzler,Sokolov2012,Meroz15}. 
Sub-diffusion caused by cytoplasmic crowding or temporal binding could obstruct the movement of the organelle \cite{Saxton,Saxton1}.  Motors which drive the organelle could cycle unproductively at junctions of three or more filaments leading to sub-diffusive behavior \cite{ Scholz}. 
However, sub-diffusion also can be biologically beneficial since it restricts organelles from moving far away from their microtubules.
Thermally driven sub-diffusion can be attributed to visco-elastic properties of the cytoplasm \cite{Weitz}, as observed for the thermal motion of colloidal tracer particles in entangled actin filament networks \cite{Weitz2004}. Mathematically it can be modeled using the fractional Langevin equation which fulfills the fluctuation-dissipation relation (FDR) \cite{Lutz,KouXie,Goychuk}. In the context of intracellular transport, thermal sub-diffusive behaviour  of non-interacting cargo-motor complexes was studied in Refs. \cite{Bouzat,GKM1,GKM2,Nam,Klein2014}. On the other hand, non-thermal sub-diffusion can be modelled as the Langevin equation with an external fractional Gaussian noise (fGn) which does not fulfill the FDR \cite{Mandelbrot,Chechkin,Waigh14}.
Previously it has been used to decribe the dissociation dynamics of biopolymers from a bound state, particle sub-diffusion under molecular crowding conditions and bulk chemical reactions of larger particles under super-dense conditions \cite{Weiss,FGN,FGN1,FGN2}. 

In the long-time limit with the lipid droplet experiments, the TAMSDs become super-diffusive: TAMSD $\sim \Delta^{\alpha}$ with $1<\alpha \le 2$ where $\alpha=2$ corresponds to ballistic motion \cite{Kenwright}. Although it is clear that organelle transport is driven by molecular motors, 
there could be various reasons for sub-ballistic and super-diffusive behaviour in living cells. The highly viscous, dissipative environment could be one of them \cite{GKM2}.
Transient super-diffusion was suggested to be caused by geometry in driven crowded systems \cite{Benichou13}, fluctuations \cite{Bruno}, competition of opposite polarity motors \cite{Klein2014,Bouzat}, heterogeneity of the substrate \cite{Sancho,Kafri}, the motion of microtubules \cite{Kulic} or asymmetry in attachment and detachment rates \cite{Hafner}. In the later model it was assumed that the motor remains attached, but stationary, to the filament during the pause state, or that it detaches, but stays immobile, at the detachment point until it attaches again. 

To understand organelle motion, various models of motor protein dynamics and those of cargo-motor complexes were developed \cite{Julicher,Kolomeisky,KKBL,Klumpp,Lipowsky,Bouzat,Klein2014,Bruno,Hafner,GKM1, GKM2,Nam}. Several of them incorporate sub-diffusive and/or super-diffusive behaviour for the cargo-motor complexes \cite{Bouzat,Klein2014,Bruno,Hafner,GKM1, GKM2,Nam}. All these models assume Markovian attachment and detachment rates of motor proteins from filaments. We suggest that non-Markovian kinetics of motors could be the reason for the cargo-motor behaviour observed in our experiments \cite{Kenwright}. 
We introduce the effective detachment rate for motors which is not constant as it is commonly used for modelling 
(see also the Markovian rate model defined below), but inversely proportional to the time interval since the last attachment event.  Recently we observed the non-Markovian detachment rate experimentally in the dynamics of cargo-motors complexes in eukaryotic cells \cite{SUBMITTED}.
One of the possible origins of the non-Markovian rate is based on the natural assumption of a heterogeneous population of motors. Cargo movement frequently involves a mixed population of motors with diverse properties of speed, detachment rate, etc. Therefore the detachment rate for a motor should be thought of as describing the effective rate of many different types of motors. The non-Markovian unbinding rate corresponds to a power-law distribution of running times of motors walking along filaments \cite{Mendez,Fedotov,Fedotov1}. Power-law distributions of running times differ from the usually assumed exponential (Poisson) distribution, which follows for a constant unbinding rate \cite{Kolomeisky,KKBL,Lipowsky}. Recently power-law distributions of running times have been observed in experiments for molecular motors in living cells \cite{Granick}. We also base our assumption of non-Markovian kinetics on the mounting evidence of anomalous internal dynamics of proteins \cite{Meroz,Reuveni}. Non-Markovian internal kinetics have been discovered in a range of enzymes based on single molecule experiments \cite{Min,Chen,Xie}. Reaction kinetics of polymers also demonstrate non-Markovian effects \cite{Guerin}. Previously, a model which incorporates some non-Markovian effects in terms of the memory in time of the averaged force acting on the cargo was considered in Ref.\ \cite{Bouzant2016}.

The proposed non-Markovian model is successful in simultaneously describing the time averaged mean squared displacement, the mean first passage time of trajectories and the multiple peaks observed in the distributions of cargo velocities. Based on the analysis of our model, we find that non-Markovian motor kinetics could be biologically beneficial by allowing a cargo to reach from one microtubule to another with much higher probability, and therefore promoting active transport. We also show that the non-Markovian kinetics could increase the average distance that the cargoes travel when microtubules are blocked by another filament, so promoting long range transport.

\section*{Results and Discussion}

The individual lipid droplet tracks were obtained using high speed bright-field microscopy \cite{Kenwright}. Combined with sophisticated tracking software \cite{Rogers2007}, it provided nanometre resolution of organelle positions at sub-millisecond time scales in live cells. The motion was imaged at 10 000 frames per second. Details of experimental settings and methods are described in Ref.\ \cite{Kenwright}. In the experiment, 2D projections of 3D trajectories were recorded. Therefore, in theoretical modelling we ignore the third dimension. However, the 2D approximation of trajectories could be well justified. Firstly, the microtubules effectively reduce the dimensionality. Secondly, if trajectories had big z-components, they would quickly go out of the focal plane of the microscope and stop being recorded. In this way only trajectories that were close to 2D were recorded. In general, resolving the third dimension is an open experimental and theoretical problem. Some spurious artifacts caused by 3D projection into 2D which transiently mimic superdiffusion in the projected coordinate were discussed in Refs.\ \cite{Proj1, Proj2}.

\begin{figure}
\centering
\vspace{0pt}
\hspace{0pt}
\includegraphics[scale=0.7]{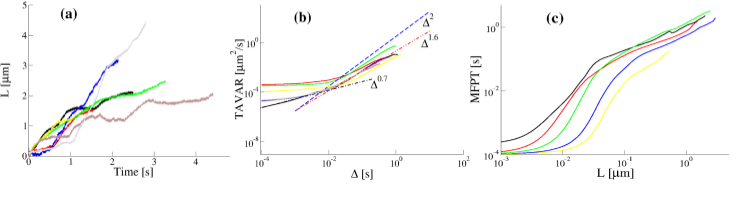}
\caption{\label{FIG1} (a) Distance $L(t)=\sqrt{x^2(t)+y^2(t)}$ traveled by single lipid droplet cargoes as a function of time, extracted from experimental trajectories \cite{Kenwright}. (b) Time averaged variance (TAVAR) for single experimental trajectories of the lipid droplets as a function of time interval $\Delta$. (c) The mean first passage time (MFPT) as a function the displacement $L$ for single experimental trajectories.}
\end{figure}

It was observed in 
experiments \cite{Kenwright} that organelles move along microtubules in straight or slightly curved trajectories. 
Several sample trajectories are shown in Fig.\ \ref{FIG1} (a). 
Since vesicles were found to move predominately in one direction towards the cell centre, we conjectured that only dynein motors were engaged and there was no tug-of-war between motors of opposite polarity. We observed steps of $8$ and $18$ nm typical for dynein motors in the cargo trajectories (Fig A and B in \nameref{S1_File}). It is not clear how many motors were engaged simultaneously. The number of working motors is difficult to determine even {\it in vitro} \cite{Ajay}. Therefore, in our modelling we pursue the mean-field approach assuming that the cargo is attached to an effective motor with effective detachment and attachment rates (see below). 

{\it Anomalous transport properties.} To analyse experimental trajectories we calculated the time averaged mean displacement (TAMD, Eq.\ (\ref{TAMEAN})), the mean squared displacement (TAMSD, Eq.\ (\ref{TAMSD})) and the time averaged variance (TAVAR, the time averaged mean squared displacement corrected for directed motion, Eq.\ (\ref{TAVAR})) of individual trajectories. The TAMD for experimental trajectories of the lipid droplets grow linearly as a function of time interval $\Delta$ (Fig C in \nameref{S1_File})
 and therefore the TAMSD grows ballistically, $\text{TAMSD}(\Delta) \sim \Delta^{2}$ for approximately $\Delta>0.02$ s (Fig D (a) in \nameref{S1_File}). Therefore, instead of TAMSD we calculated the TAVAR of a single trajectory (Fig.\ \ref{FIG1} (b)) and found a power-law dependence with the anomalous exponent $\alpha$:
\begin{equation}
\text{TAVAR}(\Delta) \sim \Delta^{\alpha}.
\end{equation}
TAVAR shows a clear transition from sub-diffusion $0<\alpha<1$ at short time scales with the exponent $\alpha \simeq 0.7$
to super-diffusive motion $1<\alpha<2$ at large times with the average exponent $\alpha \simeq 1.6$. For different experimental trajectories anomalous exponents of experimental trajectories 
were found to be in the range from $\alpha=1.35$ to $\alpha=1.77$. We suggest that this super-diffusive behaviour is asymptotic and biologically beneficial compared to normal diffusion ($\alpha=1$). Super-diffusion helps lipid droplets to overcome obstruction of motion caused by sub-diffusion, e.g. \cite{Reverey}. The transition from sub- to super-diffusion occurs at approximately a $20$ ms time interval and was reported in Ref.\ \cite{Kenwright}. The TAVARs of some trajectories saturates at small time scales (see Fig.\ \ref{FIG1} (b)). We show that this small time saturation can be explained by the measurement errors in recording of trajectories (see the discussion of Fig.\ \ref{FIG5} (a), (b) below).  

{\it Mean first passage time.} To analyse the experimental trajectories further we calculated the mean first passage time (MFPT) for each organelle to reach a certain distance $L$ for the first time. Figure \ref{FIG1} (c) shows a considerable spread of the MFPTs among the trajectories and a transition from saturated to fast growth for small $L$ to almost linear growth at longer distances. The time which corresponds to this transition ($t\sim0.1$ s) is significantly different from the crossover time between sub- and super-diffusive behaviour of the TAVAR and the TAMSD, at approximately $20$ ms. The saturation behaviour at short times and fast growth of the MFPT for small $L$ are captured by measurement errors included in our model (see the Discussion below).

\begin{figure}
\centering
\includegraphics[width=2.7in]{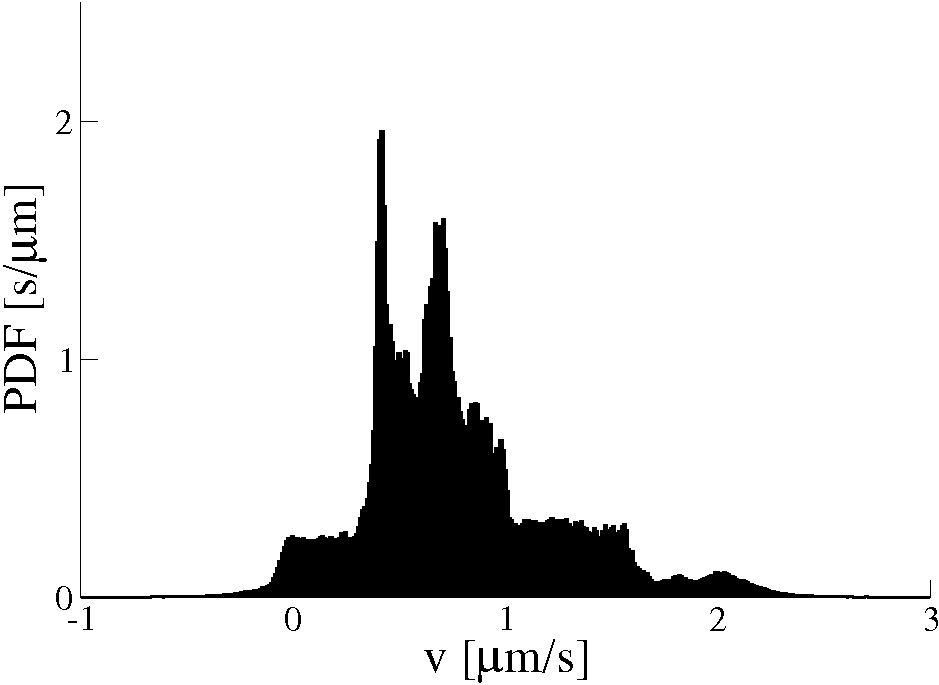}
\caption{Probability distribution function (PDF) of velocities along one experimental trajectory of a lipid cargo.}
\label{FIG2}
\end{figure}

{\it Observed cargo velocities.} Experimental trajectories have been analysed by extracting velocities of the cargoes. We calculated the distribution of average velocities (see Methods for definition). The distributions of $v$ in Fig.\ \ref{FIG2} have several peaks corresponding to average velocities of different parts of the trajectory. 
The velocity of the cargo in one trajectory varies from $0$ to $2$ $\mu$m/s. We checked that this method accurately estimates velocities of a cargo (Fig E and F in \nameref{S1_File}). More examples of distributions of cargo velocities are shown in Fig.\ G in \nameref{S1_File}. Peaks in the distributions which correspond to different cargo velocities confirm the method.

\subsection*{Results of mathematical modelling}

\begin{figure*}
\centering
\includegraphics[width=2.7in]{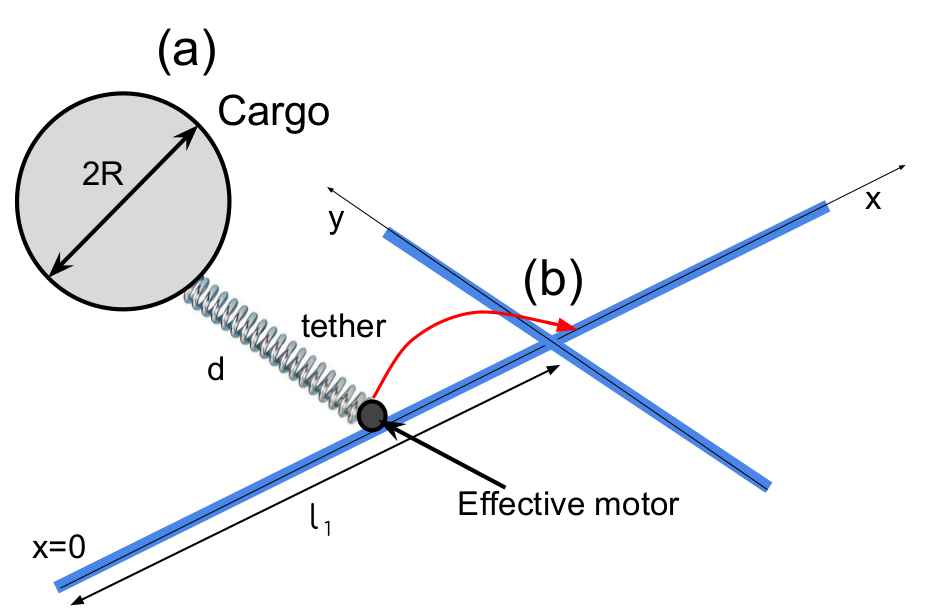}
\includegraphics[width=2.5in]{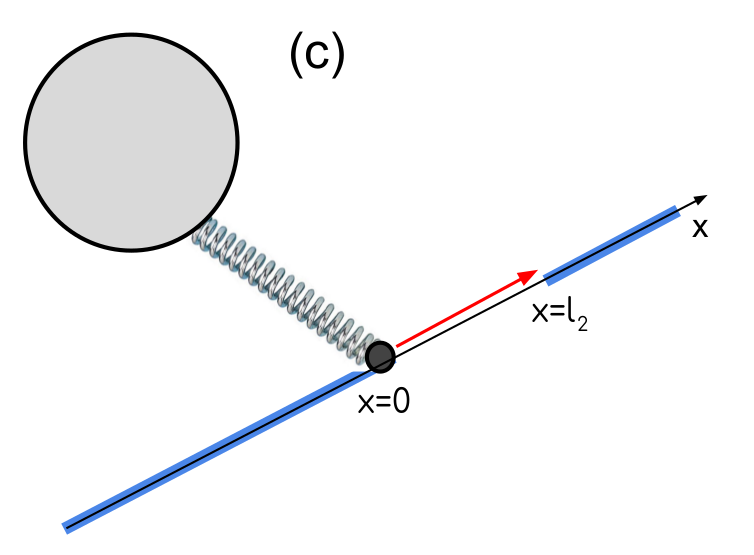}
\caption{Schematic illustration of the set-up used in simulations. (a) The cargo of radius $R$ 
moves along the filament driven by the effective motor, which is attached to the cargo by the elastic tether of length $d$. 
The microtubule is oriented along the $x$-axis. (b) Another microtubule located at distance $l_1$ from $x=0$ and 
oriented along the $y$-axis perpendicular to the $x$-axis, blocking the movement of the cargo. (c) The motor-cargo complex 
located at the end of one filament ($x=0$) is targeted to reach another filament at distance $x=l_2$.}
\label{FIG3}
\end{figure*}

\begin{figure*}
\centering
\includegraphics[width=2.5in]{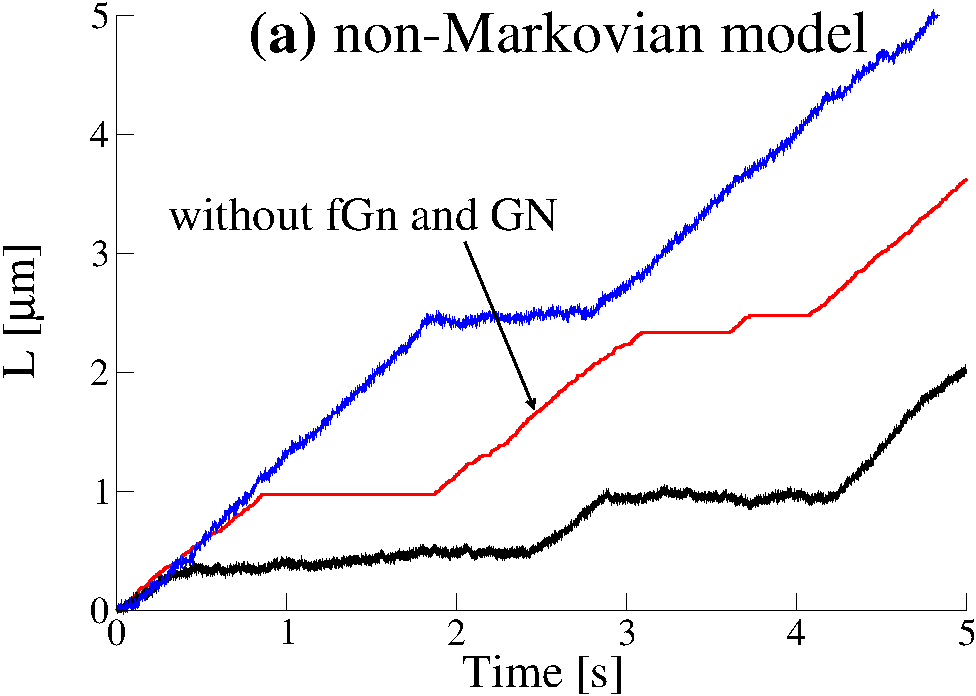}
\includegraphics[width=2.5in]{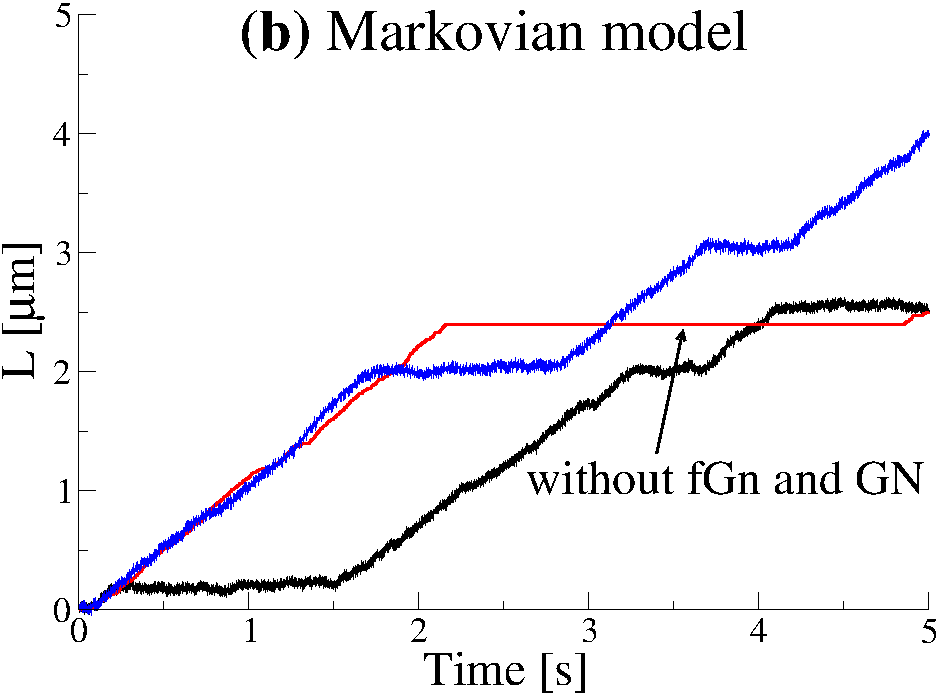}
\caption{(a) Distance $L(t)=\sqrt{x^2(t)+y^2(t)}$ travelled by cargoes as a function of time extracted from trajectories generated by the non-Markovian rate model with the parameters $\mathbb{T}_{a}=1$ s$^{-1}$, $\mu=1.4$, $\tau_d=1$ s (Eq.\ (\ref{Td1})) and (b) for the Markovian rate model. Other parameters are given in the Methods. Trajectories were generated with fractional Gaussian noise (fGn) and Gaussian noise (GN). One trajectory for each model is shown without fGn and GN for comparison.}
\label{FIG4}
\end{figure*}

To explain the experimental findings, we modelled the dynamics of a cargo complex, which is moving along a microtubule. For simplicity we do not consider the heterogeneity of the cell cytoplasm. 
We propose a mean-field approach where we assume the cargo is driven by an effective motor. Below we discuss how such a mean-field description could arise from the dynamics of multiple motors. A schematic illustration of the set-up used in simulations is shown in Fig.\ \ref{FIG3} (a). 
To better understand the implications of non-Markovian motor kinetics, we constructed non-Markovian and Markovian rate models (see Methods for definition and parameters) and compare them. Several sample trajectories are shown in Fig.\ \ref{FIG4}. 
In the non-Markovian rate model we consider a Markov attachment rate and non-Markovian detachment rate. The second model involves a Markov rate for the cargo detachment and attachment from and to the microtubule. Our numerical results suggest that contrary to the Markovian rate model, the non-Markovian rate model is able to describe asymptotic super-diffusive cargo transport. An investigation of the dynamics of a group of interacting motors and the further impact of the cytoskeletal network morphology on the intracellular transport 
\cite{Korabel,Hafner} will be given in a future study. 

\begin{figure*}
\centering
\includegraphics[width=2.5in]{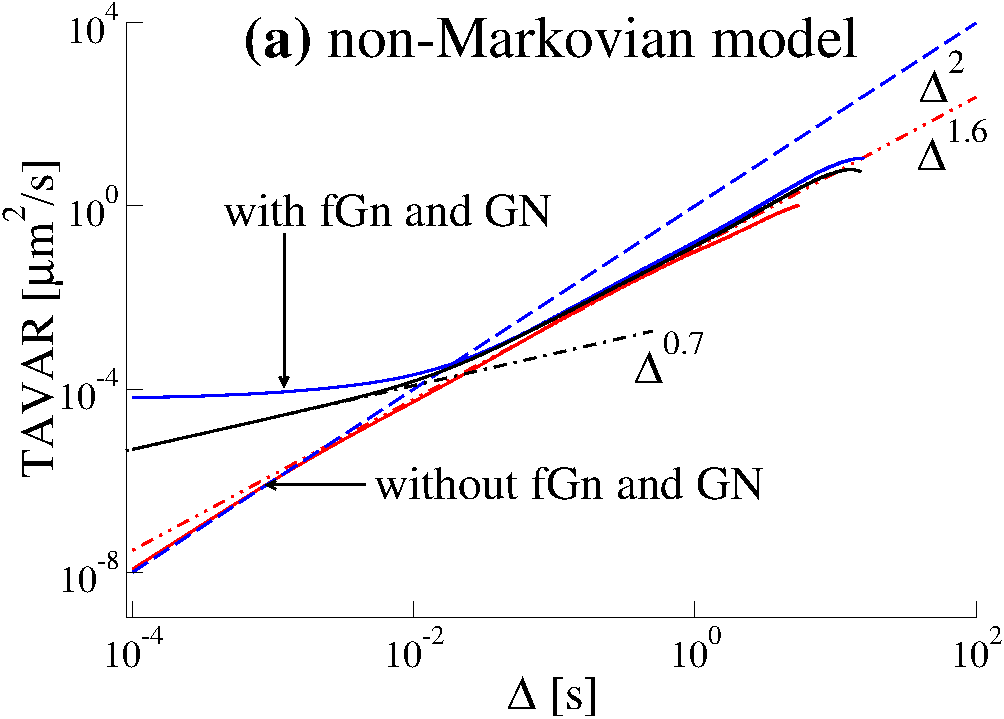}
\includegraphics[width=2.5in]{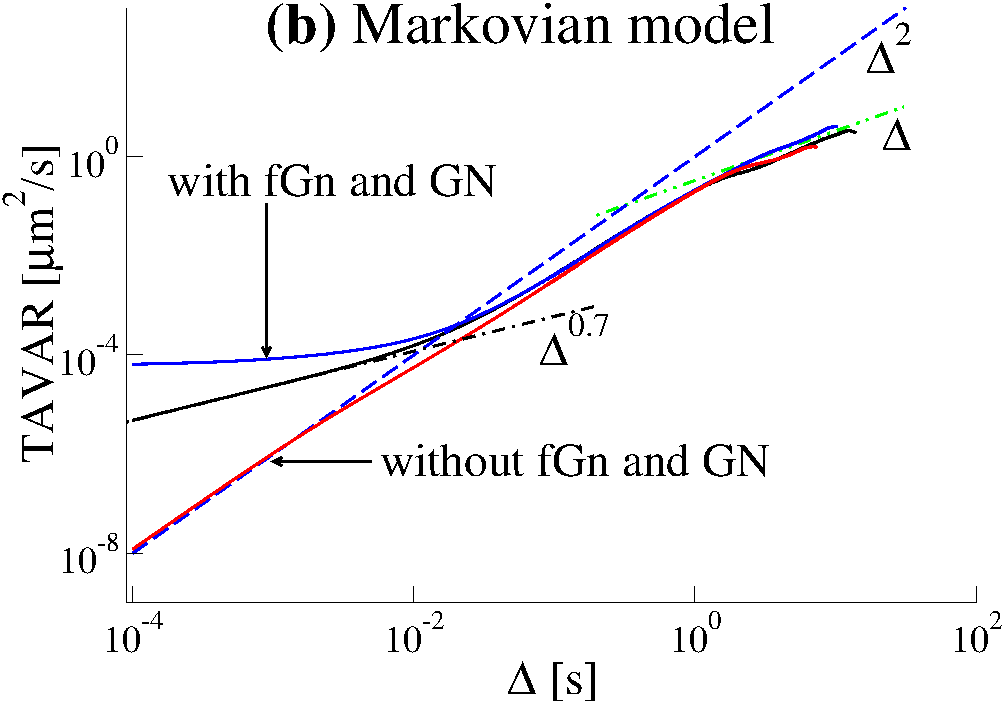}
\caption{(a) TAVAR for trajectories generated by the non-Markovian rate model and (b) by the Markovian rate model. Parameters of the models were the same as in Fig.\ \ref{FIG4}. 
Power-law scaling trends are shown with $\sim \Delta^{0.7}$ for sub-diffusion (black dashed-dotted line), $\Delta^{1.6}$ for super-diffusion (red dashed-double dotted lines), $\Delta$ for normal diffusion (green dashed-double dotted lines) and $\Delta^2$ for ballistic motion (blue dashed line).}
\label{FIG5}
\end{figure*}

\begin{figure*}
\centering
\includegraphics[width=2.5in]{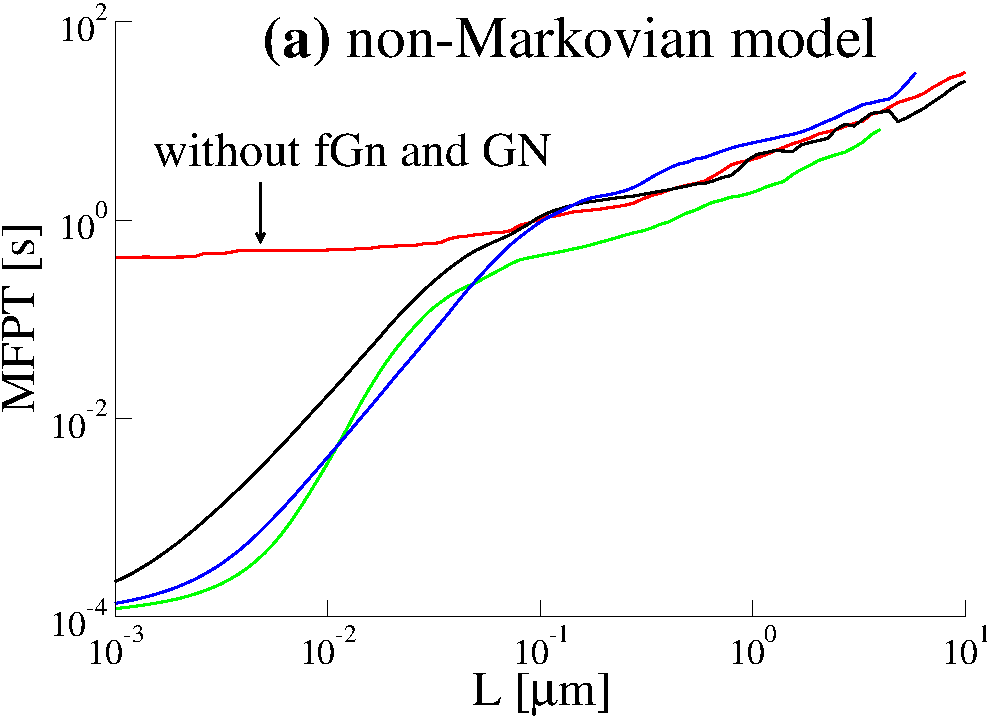}
\includegraphics[width=2.5in]{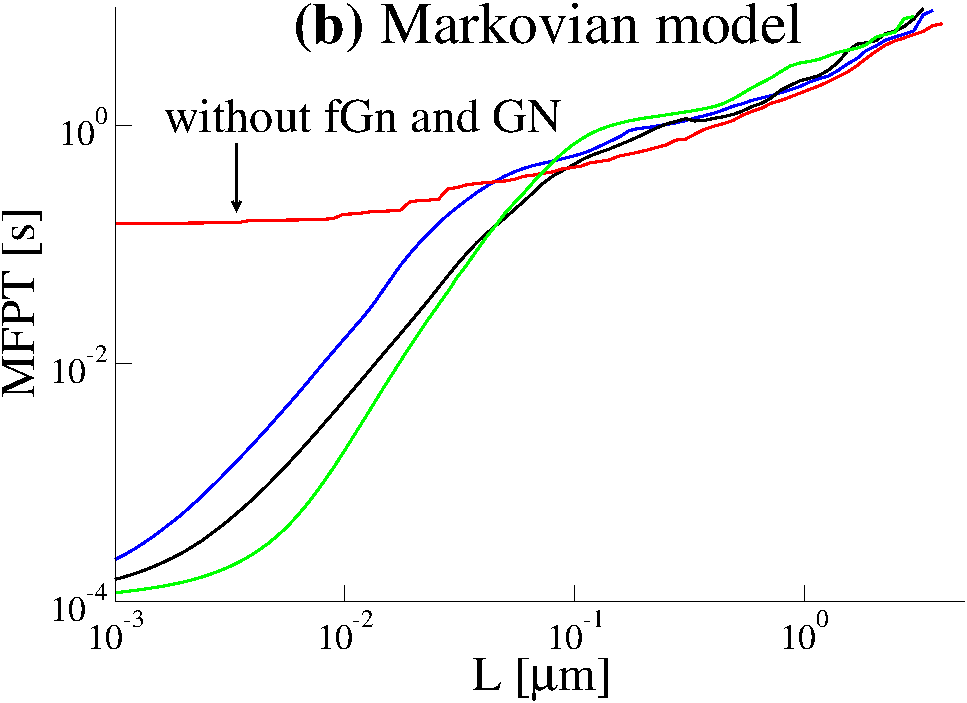}
\caption{(a) The mean first passage time (MFPT) as a function the displacement $L$ for single trajectories generated by 
the non-Markovian rate model and (b) the Markovian rate model. Parameters of the models were the same as in Fig.\ \ref{FIG4}. 
All the MFPTs have similar behaviour. For trajectories generated by models without fGn and GN, the MFPT saturates at small distances.}
\label{FIG6}
\end{figure*}

{\it The non-Markovian model reproduces super-diffusive behaviour.}
To explain the transition from sub-diffusive to super-diffusive behaviour observed in the experiment, we proposed a model 
which involves a run-length dependent non-Markovian detachment rate of the motor from the microtubule instead of a more commonly used Markovian detachment rate. The run-length dependent non-Markovian detachment rate is defined by inverse functions of the time interval $\tau$ between 
two detachment/attachment events (this time interval is called {\it a run}):
\begin{equation}
\mathbb{T}_{d}(\tau)=\frac{\mu}{\tau_d+\tau}  \exp(F/F_d),
\label{Td1}
\end{equation}
where $F$ is the absolute value of the load force acting on the motor, $F_d$ is the detachment force, $1<\mu<2$ is the anomalous exponent and $\tau_d$ is the characteristic time scale. Since Eq.\ (\ref{Td1}) is not a constant but depends on $\tau$, the detachment rate is non-Markovian. Notice that the non-Markovian detachment rate Eq.\ (\ref{Td1}) is different from an inhomogeneous Poisson process since it depends on the interval $\tau$ between two detachment/attachment events, while for an inhomogeneous Poisson process the rate depends on the time from the beginning of observation.
The switching dynamics from and to the microtubule in our model 
is similar to a L\'evy walk (LW) model interrupted by rests \cite{Sokolov} so that 
our model has super-diffusive behaviour in TAVAR for the exponent of the detachment rate $1<\mu<2$ in Eq.\ (\ref{Td1}) 
\begin{equation}
\text{TAVAR}(\Delta) \sim \Delta^{3-\mu}.
\end{equation}
We note, however, that in the standard sub-ballistic super-diffusive LW model the
time average mean displacement (TAMD) is zero, whereas in our model it grows linearly with time. 
The TAMSDs in our model are ballistic. To distinguish the motion which we study from the standard super-diffusion which is
commonly associated with the super-linear sub-ballistic behaviour of MSDs or TAMSDs, we shall call it {\it super-diffusion
in TAVAR}. We also note that super-diffusion in TAVARs is equivalent to the super-diffusion in TAMSDs if the trajectories are detrended before TAMSDs are calculated. The combination with fGn, which is independent of the non-Markovian detachment rate, leads to sub-diffusion at small times. We performed extensive simulations of the non-Markovian model for different parameters of the model, namely, the anomalous exponent $\mu$ and the characteristic time scale $t_d$ defined in Eq.\ (\ref{Td1}). The time scale $t_d$ controls the transition from sub-diffusive to super-diffusive behaviour and was chosen from $1$ to $0.0001$ seconds. Such a wide range of parameters might help to maintain robust transport inside living cells. The TAVARs show a clear transition from sub-diffusion to super-diffusion (Fig.\ \ref{FIG5} (a)). For $\mu=1.4$, the power-law exponents $\alpha=3-\mu=1.6$. So, the non-Markovian rate model reproduces the super-diffusive behaviour of experimental trajectories with the average anomalous exponent $\alpha=1.6$ (Fig.\ \ref{FIG1} (b)). The TAVAR clearly distinguishes 
the non-Markovian rate model from the Markovian rate model (defined below). The MFPT of numerically generated trajectories shown in Fig.\ \ref{FIG6} (a) also behaves similar to the MFPT of the experimental trajectories (Fig.\ \ref{FIG1} (c)).

To our knowledge there exist only a few mathematical models for
super-diffusive intracellular transport in the literature. On of them, Ref.\ \cite{Bruno}, modeled the super-diffusive behaviour as a correlated motion. In Refs.\ \cite{Klein2014,Bouzat,GKM2}, transient super-diffusion occur due to highly viscous, dissipative environment. The nature of super-diffusive behaviour in our model which is introduced below is the non-Markovian effective detachment rate of the motors from a filament experimentally observed in Ref. \cite{SUBMITTED}. 
Below we suggest several possible scenarios for how the non-Markovian detachment rate could emerge from the dynamics of multiple motors.

{\it Possible origins of the non-Markovian detachment rate.}
Cargo movement frequently involves a mixed population of dyneins
with diverse properties such as speed, detachment rate, etc.
Therefore the detachment rate $\mathbb{T}_{d}$ in the Eq. (\ref{Td1}) describes an effective rate. Below we discuss different scenarios of how the effective non-Markovian detachment rate could arise from the dynamics of multiple motors.   

\begin{figure*}
\centering
\includegraphics[width=2.5in]{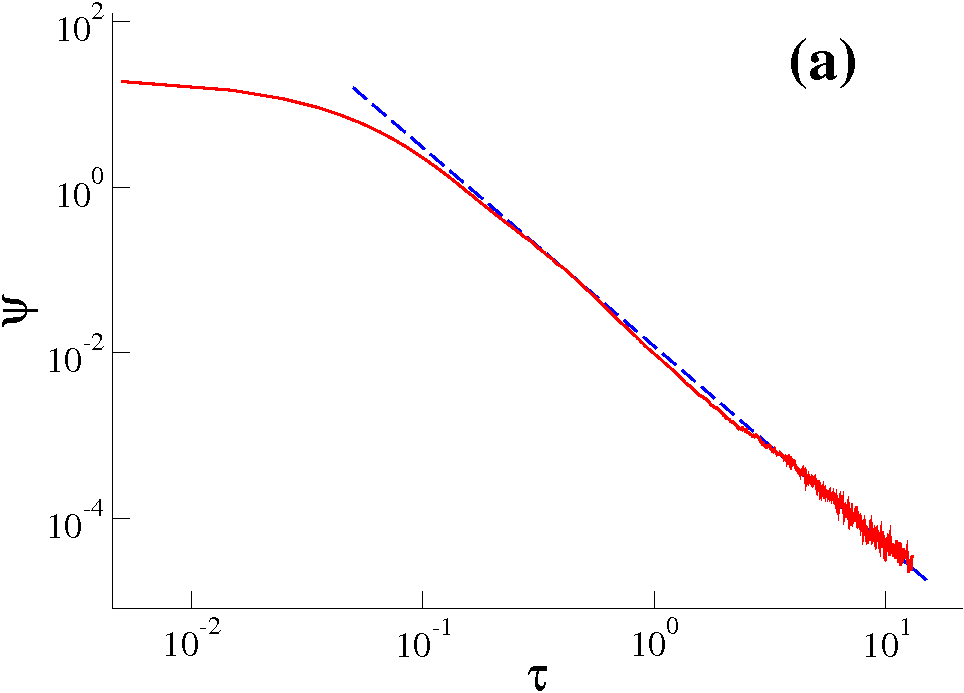}
\includegraphics[width=2.5in]{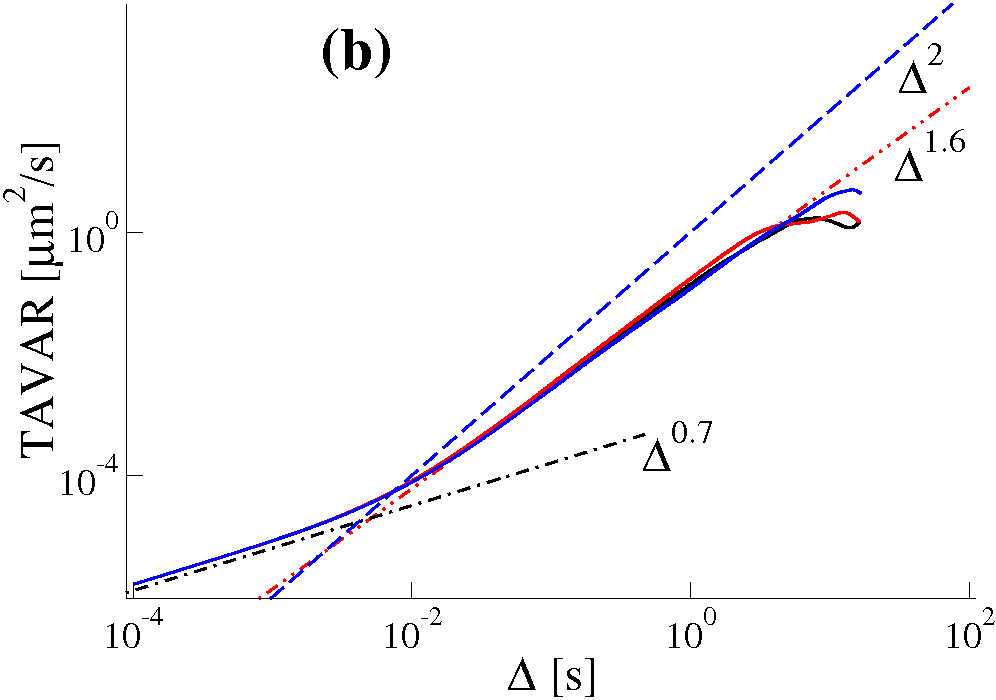}
\caption{(a) The probability density function (pdf) of residence times of a cargo on a microtubule, $\psi(\tau)$, for an ensemble of $N=6$ motors (solid curve) calculated using Eq.\ (\ref{psi}) with a particular set of $p_i$ and $\lambda_i$. We chose $p_i$ such that $\sum_i p_i =1$. The dashed curve is the power-law fit 
$\psi(\tau) \sim A \tau^{-\mu-1}$ with $\mu=1.4$ and $A=0.012$. This power-law pdf corresponds to the rate $\mathbb{T}_{d}(\tau)\sim \mu/\tau$, which is consistent with the long time behaviour of Eq.\ (\ref{td}). Notice that 12 parameters ($p_i$ and $\lambda_i$) are parametrized by only two parameters 
$A$ and $\alpha$. (b) The time averaged variances TAVARs 
(the time averaged mean squared displacements corrected for the drift) calculated along single trajectories generated with the residence times of the cargo on the microtubule shown in (a) grow as $\Delta^{3-\mu}=\Delta^{1.6}$ (red dashed-double-dotted curve).}
\label{FIG7}
\end{figure*}

{\bf Case 1}. We assume that the cargo is attached to $N$ motors and only one motor is engaged with the microtubule. We also assume that there are $N$ types of motors such that a single motor with the probability $p_{i}$ has an exponential distribution of engagement times with a constant detachment rate $\lambda_{i}$. The motors with the higher rate $\lambda_{i}$ are detached rapidly and the motors with the lower rates will dominate in the long time limit. In what follows we explain theoretically why the decreasing rate in Eq. (\ref{Td1}) is a good approximation and justify it in numerical simulations.
The crucial question is: what is the effective detachment rate 
$\mathbb{T}_{d}$ in this case? To find this rate, we define the 
probability density function of the cargo engagement time $\psi (\tau )$ on microtubule and the corresponding survival function $\Psi (\tau )=\int_{\tau}^{\infty} \psi (\tau ) d \tau $.
In our case:
\begin{equation}
\psi (\tau )= \sum_{i = 1}^{N} p_{i} \lambda_{i} e^{-\lambda_{i} \tau}, \; \; \; \Psi (\tau )= \sum_{i = 1}^{N} p_{i} e^{-\lambda_{i} \tau }.
\label{psi}
\end{equation}
According to the definition of the detachment rate \cite{Cox}:
\begin{equation}
\mathbb{T}_{d}(\tau)=\frac{\psi (\tau )}{\Psi (\tau )}.
\label{rate}
\end{equation}%
It follows from these formulas that $\mathbb{T}_{d}(\tau)$
 is a decreasing function of the running (engagement) time $\tau $. At $\tau =0$, the effective detachment rate takes the maximum value and this is just the mean value
 $\mathbb{T}_{d}(0)= \sum_{i = 1}^{N} p_{i} \lambda_{i}.$
In the long time limit the detachment rate $\mathbb{T}_{d}(\tau)$ tends to the smallest value of $\lambda_{i}$.
 One can choose the parameters $N,$ $p_{i}$ and $\lambda_{i}$ to have a perfect fit with the formula:
\begin{equation}
\mathbb{T}_{d}(\tau)=\frac{\mu}{\tau_d+\tau},
\label{td}
\end{equation}
within any time interval. For simplicity we consider $F=0$ here. 
The exponential term $\exp(F/F_d)$ in Eq. (\ref{Td1})
does not depend on $\tau$ and therefore plays the role of a scaling factor. In Fig.\ \ref{FIG7} (a) the probability density function (pdf) of residence times of a cargo on a microtubule $\psi(\tau)$ generated using $N=6$ motors with a heterogeneous set of exponential detachment rates is shown. 
The residence time pdf is well approximated by the power law $\psi(\tau) \sim A \tau^{-\mu-1}$ with $\mu=1.4$ and $A=0.012$ within a broad time interval. This power-law pdf corresponds to the rate $\mathbb{T}_{d}(\tau)\sim \mu/\tau$, which is consistent 
with the long time behaviour of Eq.\ (\ref{td}). The TAVARs of trajectories calculated with these residence times grow as $\Delta^{3-\mu}=\Delta^{1.6}$ (see Fig.\ \ref{FIG7} (b)) 
similar to TAVARs of experimental trajectories (Fig.\ \ref{FIG1} (b)) and to TAVARs of trajectories generated by the non-Makovian rate model (Fig.\ \ref{FIG5} (a)). We note that the trajectories of the model with $N=6$ motors consist of several hundreds of 
detachment-attachment events therefore the use of the time average quantities is justified in our mean-field approach. In our numerical simulation we generate trajectories as follows: at t=0, we select a motor with the detachment rate $\lambda_i$ with the probability $p_i$. Then we chose a random number $T$ drawn 
from exponential the distribution with the mean $1/\lambda_i$. During this time the cargo moves with the velocity $v$
a distance $v T$. After this time the cargo detaches from the microtubule and rests during the random time $T^{\ast}$
drawn from another exponential distribution with the rate $\mathbb{T}_{a}=1$/s (see Methods). After this period of time the attachment process repeats. 

\begin{figure*}
\centering
\includegraphics[width=2.5in]{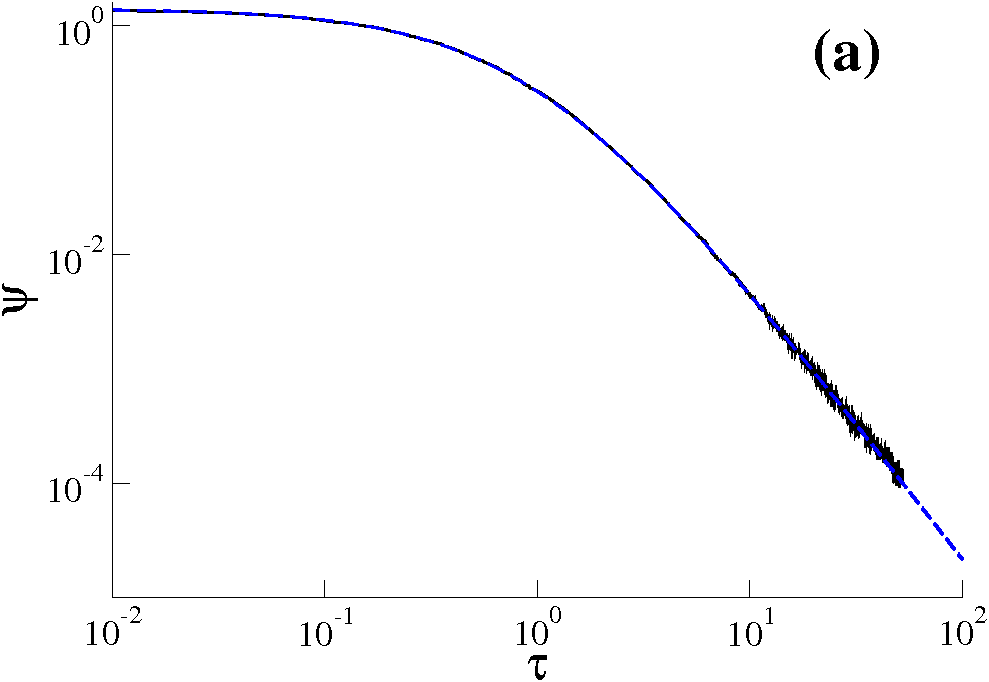}
\includegraphics[width=2.5in]{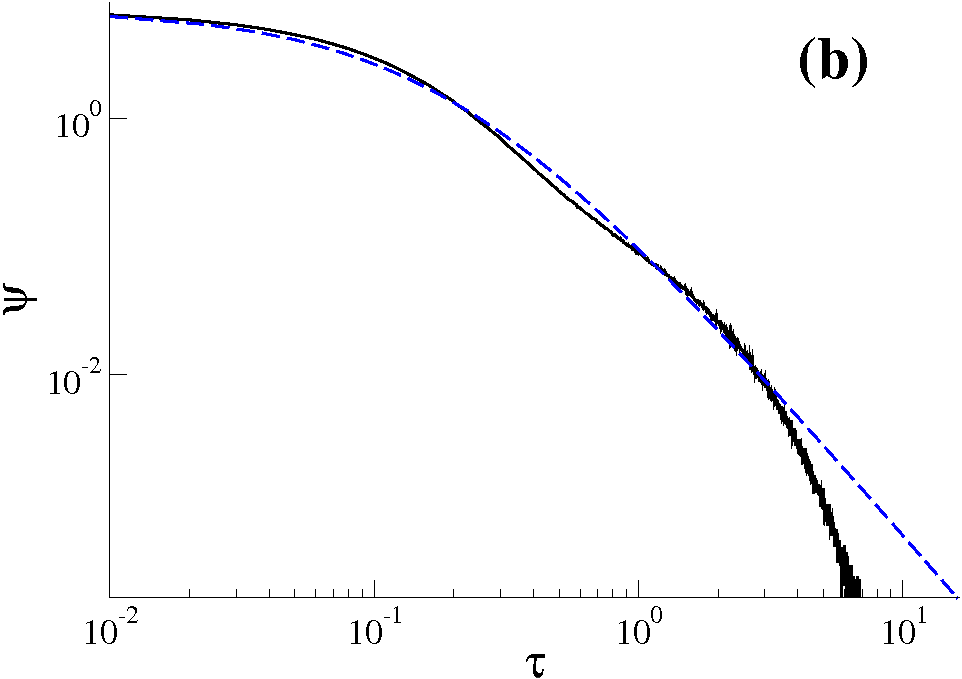}
\caption{(a) The probability density function $\psi(\tau)$ (black solid curve) of resident times of cargo on a microtubule  
for a heterogeneous ensemble of motors with $\lambda$ distributed with the gamma density Eq.\ (\ref{gamma}) with $\mu=1.4$  and $\tau_d=1$ s (Case 2). The blue dashed curve is given by Eq. (\ref{rate3}). (b) The probability density function $\psi(\tau)$ of resident times of cargo on a microtubule (black solid curve) 
for $N=3$ cooperative motors (Case 3). Other parameters are given in Ref.\ \cite{KlumppLipowsky}. The blue dashed curve is given 
by Eq. (\ref{rate3}) with $\mu=1.4$  and $\tau_d=0.18$ s.}
\label{FIG8}
\end{figure*}

{\bf Case 2}. 
To obtain the detachment rate $\mathbb{T}_{d}(\tau)$ as
the inverse function of the engagement time, $\tau$  Eq.\ (\ref{td}), we extend the formula Eq.\ (\ref{psi}) by allowing $\lambda_{i}$  to be continuously distributed
with the gamma density  \cite{Feller,Singer}:
\begin{equation}
p(\lambda)=\frac{\tau_{d}^{\mu} \lambda^{\mu -1}e^{-\tau_{d}\lambda}}{\Gamma (\mu )}.
\label{gamma}
\end{equation}%
This function gives the proportion of motors with the detachment rate $\lambda$ in the population of motors. The gamma distribution has a single peak ($\mu>1$) which decays exponentially at long times and has a negligible probability 
for instantaneous detachment; both assumptions seem physically reasonable. Therefore, the probability density function of the cargo engagement time $\psi (\tau )$ on microtubule and the survival function $\Psi (\tau )$ are:
\begin{equation}
\psi (\tau )=\int_{0}^{\infty} \lambda e^{-\lambda \tau} p(\lambda) d\lambda= 
  \frac{\mu}{\tau_{d}+\tau}
  \left ( \frac{\tau_{d}}{\tau_{d} + \tau }\right )^{\mu},
\label{rate3}
\end{equation}%
and
\begin{equation}
\Psi (\tau )=\int_{0}^{\infty} e^{-\lambda \tau} p(\lambda) d\lambda= \left ( \frac{\tau_{d}}{\tau_{d} + \tau }\right )^{\mu}.
\label{rate4}
\end{equation}%
Substitution of Eq.\ (\ref{rate3}) and Eq.\ (\ref{rate4}) into Eq.\ (\ref{rate}) leads to Eq.\ (\ref{td}). The numerically obtained residence time distribution shown in Fig.\  \ref{FIG8} (a) follows Eq.\ (\ref{rate3}). 
Therefore, the justification of the non-Markovian rate (\ref{td}) is based on the natural assumption of the heterogeneous population of motors. 
As it follows from Eq.\ (\ref{rate4}), the unbinding rate Eq.\ (\ref{rate}) corresponds to a power-law distribution of running times of motors walking along filaments \cite{Mendez,Fedotov,Fedotov1}.  
 
{\bf Case 3.} Another scenario is that the cargo is pulled by up to $N$ motors in a cooperative manner (the Klumpp-Lipowsky model \cite{KlumppLipowsky}). Then the binding time probability density function $\psi (\tau )$ is the linear combination of exponential functions as in Case 1 (see Eq.\ (\ref{psi}))
\begin{equation}
\psi (\tau )= \sum_{i = 1}^{N} Res (-z_{i}) e^{-z_{i} \tau},
\label{psi2}
\end{equation}
where $-z_{i}$ are the poles of the Laplace transform of $\psi (\tau )$ and $Res (-z_{i})$ are the corresponding residues. These parameters are the functions of the binding and unbinding rates. 
Although the parameters $-z_{i}$ are real and negative, the essential difference between Eq.\ (\ref{psi}) and Eq.\ (\ref{psi2}) is that $Res (-z_{i})/ z_{i}$  are no longer probabilities as $p_{i}$ in Eq.\ (\ref{psi}) and they 
do not all belong to the interval $[0,1]$. By using Eq.\ (\ref{psi2}) together with Eq.\ (\ref{rate}) one can show 
that the effective detachment rate $\mathbb{T}_{d}(\tau)$ is again a decreasing function of the running time $\tau$.
Our numerical experiments for $N=3$ (see Fig.\  \ref{FIG8} (b)) demonstrate a good agreement with formula (\ref{td}) in a broad time interval. However, eventually the dynamics of this model is Markovian (due to the exponential tail of the probability density of the residence times in Fig.\  \ref{FIG8} (b))) and therefore it can not explain the long-time super-diffusive behaviour of cargoes observed experimentally.  

We note that only Case 3 involves multiple motors. However, it does not reproduce the super-diffusive behaviour of cargoes observed experimentally. In Case 1 and 2 it is assumed that only one motor is engaged from the ensemble of motors. We will consider the effect of relaxing this assumption in the future work. Further generalizations of these three cases are possible that could involve a range of distributions and different mechanisms of attachment and detachment. The advantage of our non-Markovian 
rate model Eq.\ (\ref{td}) is that it involves only two parameters: the non-dimensional anomalous exponent $\mu$ and the time scale $\tau_d$. 

{\it The Markovian rate model leads to normal diffusion.} To insure that the non-Markovian effective detachment rate is 
the reason for super-diffusive behaviour, we investigated the Markovian rate model and find that it leads to normal diffusion.
In the Markovian rate model the binding and unbinding rates of a 
motor are constant which results in an exponential distribution for the residence times in attached and detached states. The ensemble averaged mean squared displacement (EAMSD) for this model has an asymptotic ballistic behavior \cite{Hafner}. 
The TAMSD of single trajectories is also ballistic at large time intervals (Fig.\ S7 (c)). As expected, the TAVAR grows linearly with the time interval $\Delta$, see Fig.\ \ref{FIG5} (b). The fGn leads to sub-diffusion at small times. Thus, the model demonstrates a transition from sub-diffusive to normal diffusive behaviour. Additive measurement noise leads to saturation of the TAVAR at short times also observed in TAVARs of experimental trajectories  (Fig.\ \ref{FIG1} (b)). The MFPT of numerically generated trajectories (Fig.\ \ref{FIG6} (b)) shows a good 
agreement with the MFPT of the experimental trajectories (Fig.\ \ref{FIG1} (c)). 

{\it The non-Markovian model could be beneficial in overcoming a blockage on a microtubule.} 
One can ask about the difference between the non-Markovian and Markovian rate models besides their different asymptotic transport behaviour. The transport in the Markovian rate model is asymptotically normal in TAVAR, while it is super-diffusive in TAVAR in the non-Markovian rate model. Can non-Markovian motor kinetics be biologically beneficial for intracellular transport? To address this question, we considered a cargo moving along the microtubule oriented along $y=0$. 
The microtubule was blocked by another perpendicular filament. The diameter of the blocking microtubule was $24$ nm (Fig.\ \ref{FIG3} (b)). 
In simulations the cargo was interacting with the MT through the effective motor: the motor could not make further steps once it reached the blocking MT. The only possibility for the cargo to overcome the blocker was to first detache from the microtubule and then reattach to the (possibly different but pointing in the same direction) MT. We did not consider the possibility that an effective motor switches direction by attaching to the blocking MT, which is sometimes observed experimentally. 

\begin{figure}[t]
\centering
\includegraphics[width=2.7in]{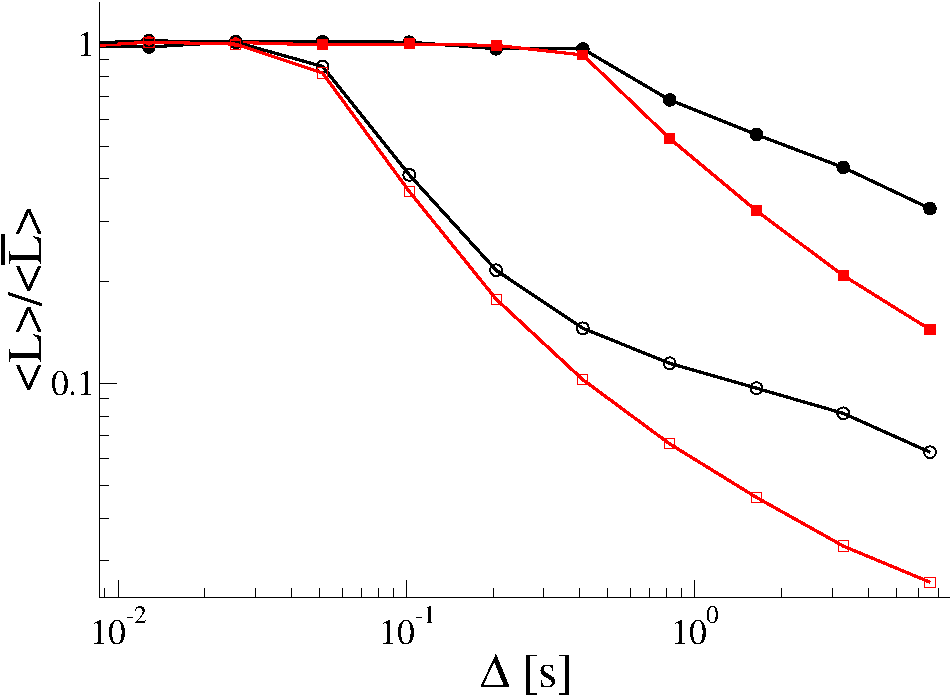}
\caption{Average distance $\left< L \right>$ the cargo travelled over a time interval $\Delta$ on the microtubule with a blocking filament located at distance $l_1=0.1$ $\mu$m (curves with open symbols) and $l_1=0.5$ $\mu$m (curves with filled symbols) from $x=0$, see Fig.\ \ref{FIG3} (b). The distance $\left< L \right>$ was normalized by the average distance $\left< \bar{L} \right>$ the cargo travelled without the blocking filament. We used $3000$ trajectories generated by the Markovian rate model (red squares) and the non-Markovian rate model (black circles) to calculate the average distances. Parameters of the models are the same as in Fig.\ \ref{FIG4}.}
\label{FIG9}
\end{figure}

The question is how to compare the Markovian and the non-Markovian rate models? To do this, we have chosen their parameter values in such a way that the average distance  by a cargo without any blockers over a fixed time period would be the same for both models. 
We took this time period to be 2 seconds which was motivated by the average duration of experimental trajectories.
For the Markovian model the load free detachment rate was fixed to $\epsilon=0.25/$s. For the non-Markovian model parameters for the detachment rate Eq.\ (\ref{Td1}) could be chosen over a wider range. For the simulations in Fig.\ \ref{FIG9} we have used  $\mu=1.4$, $\tau_d=1$ s which gives the load free detachement rate  Eq.\ (\ref{Td1}) $T_d=1.4$ s$^{-1}$. This is $5.6$ times bigger than the load free detachment rate for the Markovian model. We assumed that the only possibility for the cargo to overcome the blocker is first to detach from the microtubule and then to reattach and that the blocker is located close to the origin $x=0$. For the non-Markovian model the cargo will detach from the MT quicker than in the Markovian model because of larger detachment rate at the beginning. These factors cause a bigger distance to be travelled with blocking filaments over a fixed amount of time. We note that if the detachment rates of the Markovian and the non-Markovian model are the same or the blocker is far away from the origin $x=0$, there will be no effect.

We calculated the average distance $\left<L\right>$ travelled by the cargo over the time period $\Delta$ s normalized by the average distance $\left< \bar{L} \right>$ travelled by the cargo in the absence of the blocker. The average distances were calculated using $3000$ trajectories by sliding the time window $\Delta$ along the trajectories. Results shown in Fig.\ \ref{FIG9} suggest that the non-Markovian rate model allows cargoes to be transported longer distances for the same period of time compared to the Markovian rate model. An investigation of the influence of complex cytoskeletal network morphology will be given in a future study.

\begin{figure}
\centering
\includegraphics[width=2.7in]{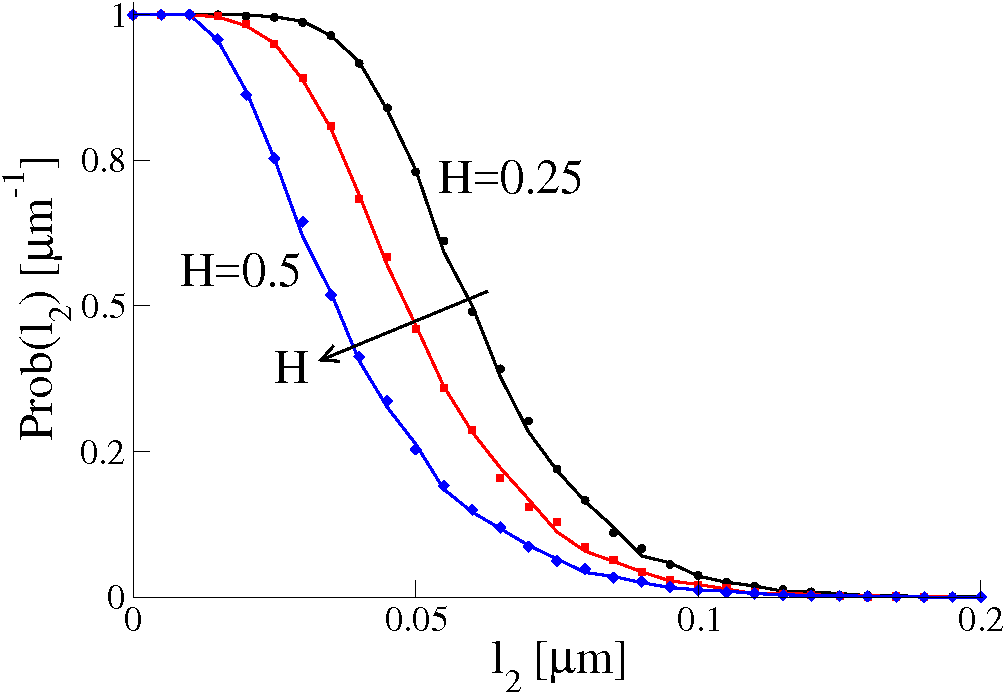}
\caption{Probability $\text{Prob}(l_2)$ for a cargo to reach a neighbouring filament located at distance $l_2$ $\mu$m (see Fig.\ \ref{FIG3} (c)) in the time interval $\Delta T=16.4$ s. 
The probability is calculated for Hurst exponents of $H=0.25$ (black), $H=0.35$ (red) and $H=0.5$ (blue) (see the arrows for the trend) using $3000$ trajectories generated by the Markovian rate model (symbols) and non-Markovian rate model (curves). Parameters of the models were the same as in Fig.\ \ref{FIG4}.}
\label{FIG10}
\end{figure}

{\it Sub-diffusion helps the cargo to reach a neighbouring filament.} 
To show this we calculated the probability for the motor-cargo complex located at $x=0$ to reach a neighbouring microtubule located at some distance $l_2$ from it, in a finite time interval $\Delta t$ (see Fig.\ \ref{FIG3} (c)). The microtubule has a diameter of $24$ nm. For simplicity we consider no force acting on the cargo as it moves between microtubules. The cargo diffused (for the fGn with the Hurst exponent of $H=0.5$) or performed sub-diffusion (for $H<0.5$) in 2D until it reached a neighbouring microtubule located at distance $l_2$ along $x$ axes. Figure \ref{FIG10} shows the probability of reaching the neighbouring microtubule as a 
function of the distance $l_2$ between filaments. 
Sub-diffusion generated by independent fGn increases the probability of reaching the target \cite{FGN}. It does not allow the cargo to drift far away and thus increases the probability of reaching another filament. 
Notice that in order to compare fractional Brownian motions with different Hurst exponents, $H$, we followed
the procedure proposed by Guigas and Weiss in Ref. [60]. Namely, we fix parameter $D^*$ and chose
the mean squared displacement of walkers for a single time step $\delta t=1$ to be equal for all simulations. 
Physically this corresponds to an equal amount of energy consumed by walkers to complete a single time step. 
With such a choice we ensure a fair comparison between fractional Brownian motions with constant $D^*$ but different $H$.

\begin{figure*}
\centering
\includegraphics[width=2.5in]{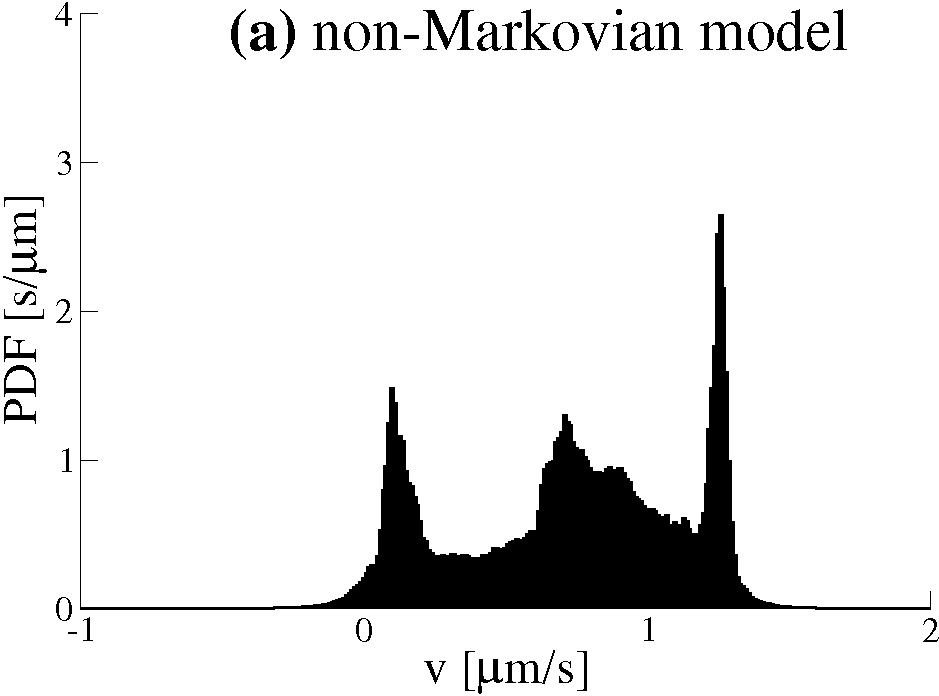}
\includegraphics[width=2.5in]{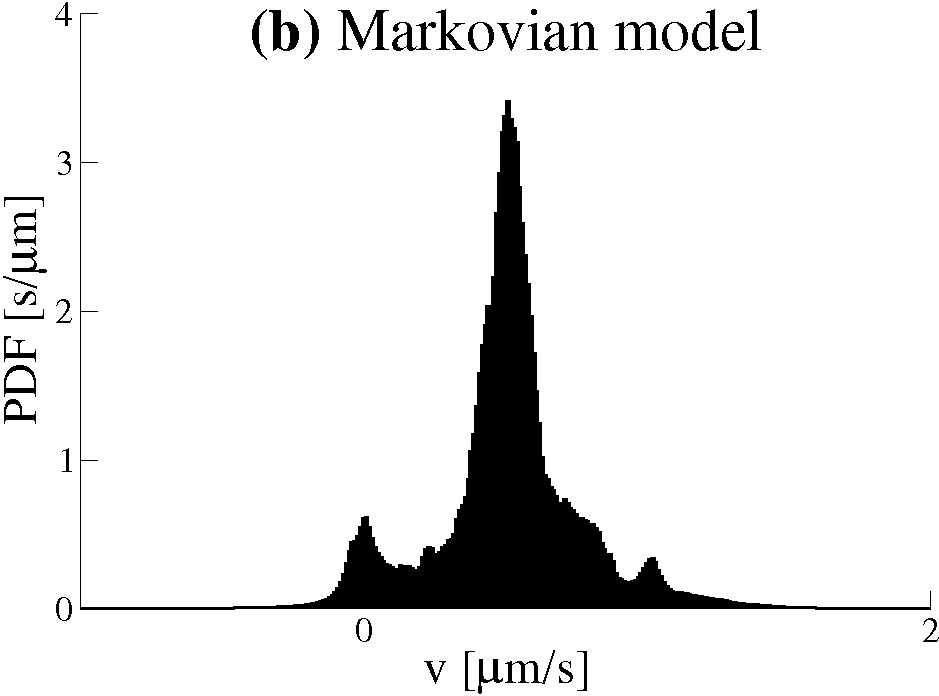}
\caption{(a) Probability distribution function (PDF) of velocities along typical trajectory generated by the non-Markovian rate model and (b) for a trajectory generated by the Markovian rate model. Parameters of the models were the same as in Fig.\ \ref{FIG4}.}
\label{FIG11}
\end{figure*}

{\it Cargo velocities.} Finally we find the cargo velocity in simulations has a value in the range of $0$ to $2$ $\mu$m/s 
(for an unloaded motor velocity $v_0=4$ $\mu$m/s) with the multiple peaks in the velocity PDF (Fig.\ \ref{FIG11} (a)), 
similar to the cargo velocities measured in the experiments (Fig.\ \ref{FIG2}). The multiple peaks in the velocity PDF are 
due to the non-Markovian nature of the model and do not require additional molecular details to be invoked to explain them e.g. different gears of the motor proteins. By contrast, the velocity PDF for the Markovian model (Fig.\ \ref{FIG11} (b)) has a few smaller peaks and a dominant peak corresponding to the average velocity of the cargo.

\section*{Methods}

\subsection*{Analysis of experimental data}
 
{\it Time averaged mean squared displacement (TAMSD).} Two-dimensional trajectories of organelles, $\mathbf{r}(t)=\{x(t),y(t)\}$, were analyzed using the TAMSD defined as \cite{Burov,Metzler}:
\begin{equation}
\text{TAMSD}(\Delta) = \frac{1}{T-\Delta} \int_{0}^{T-\Delta} \left[ \mathbf{r}(t'+\Delta) - \mathbf{r}(t') \right]^2 dt',
\label{TAMSD}
\end{equation}
where $T$ is the total duration of a trajectory and $\Delta$ is the time interval which defines the width of the window slid along the trajectory. In the experiment with liquid droplets \cite{Kenwright}, trajectories were obtained at sampling time-intervals $\delta=10^{-4}$s. The discrete version of Eq.\ (\ref{TAMSD}) reads:
\begin{equation}
\text{TAMSD}(\Delta) = \frac{\sum_{i=0}^{T/\delta-n} \left[ \mathbf{r}(i\delta+\Delta) - \mathbf{r}(i\delta) \right]^2}{T/\delta-n},
\label{TAMSD_discrete}
\end{equation}
where the averaging window is $\Delta = n\delta$. The TAMSDs grow as a power law TAMSD$=D_{\alpha} \Delta^{\alpha}$ 
where $D_{\alpha}$ is the fractional diffusion coefficient and the exponent $\alpha=1$ corresponds to Brownian diffusion, $\alpha<1$ sub-diffusion, $\alpha>1$ super-diffusion, and $\alpha=2$ 
ballistic motion. The anomalous exponent $\alpha$ can be calculated as \cite{Garini2}
\begin{equation}
\alpha = \frac{\log(\text{TAMSD})}{\log  \Delta}.
\label{alpha}
\end{equation}
The time averaged mean displacement (TAMD) along a trajectory is given by:
\begin{equation}
\text{TAMD}(\Delta) = \frac{1}{T-\Delta} \int_{0}^{T-\Delta} \left[ \mathbf{r}(t'+\Delta) - \mathbf{r}(t') \right] dt',
\label{TAMEAN}
\end{equation}
The time averaged variance is defined as \cite{FK}:
\begin{equation}
\text{TAVAR}(\Delta) = \text{TAMSD}(\Delta) - (\text{TAMD}(\Delta))^2.
\label{TAVAR}
\end{equation}
Here the TAMSD is defined in Eq.\ (\ref{TAMSD}) and the TAMD is given by Eq.\ (\ref{TAMEAN}). The definition of the TAVAR Eq.\ (\ref{TAVAR}) is equivalent to the path-wise correction for directed motion which can be realized by subtracting the mean increment from each trajectory in such a way that
the modified trajectory is constrained to begin and end at the same spatial position. The TAMSD Eq.\ (\ref{TAMSD}) of the modified trajectory then coincides with the TAVAR 
Eq.\ (\ref{TAVAR}) of the original trajectory. 
We note that other definitions of time-average variance exist. One of them uses the mean defined as $\int_{0}^{T} \mathbf{r}(t') dt'/T$ \cite{Wu}. However, this definition does not suite our problem since TAVAR for a super-diffusive trajectory defined in this way shows linear growth in $\Delta$. 

{\it Mean first passage time (MFPT).} Next we calculated the time needed to reach a point at distance $L$ along a single trajectory for the first time \cite{Redner,Chou}
\begin{equation}
T=\inf \{ \Delta: R(\Delta)=L\},
\end{equation}
where $R(\Delta)=| \mathbf{r}(t'+\Delta) - \mathbf{r}(t') |$ and $\inf$ is the infimum. The MFPT is defined as the average of $T$ i.e. MFPT$=\left< T \right>$. 

{\it Observed cargo velocities.} To extract cargo velocities from experimental trajectories we avoided manual or automatic segmentation of trajectories into segments of constant velocity, which would require using some (arbitrary) threshold. Instead, we calculated the distribution of cargo velocities defined 
as $v=(\mathbf{r}(t'+\Delta) -\mathbf{r}(t'))/\Delta$ along a trajectory. To attain better statistics, the time increments $\Delta$ was varied from its minimal value $\Delta=0.0001$ seconds to the maximal value which we set to $1/10$th of the length of a trajectory. We note that the unweighted histograms were calculated. Since we have considered only 1/10th of the length of each trajectory, the total number of velocities for different $\Delta$ did not change dramatically. We checked that the method accurately estimates velocities of a cargo (see Figs.\ F and G in in \nameref{S1_File}).

\subsection*{Stochastic models}

For the sake of simplicity, we do not address the question of the actual number of engaged motors or the heterogeneity of the cellular environment. We also ignored motor detachments from the cargo since they occur much less frequently than motor detachments from the microtubule and did not affect our calculations. When the motor detached from the microtubule, a circular cargo of radius $R$ performs two-dimensional sub-diffusion which we modelled with fractional Gaussian noise. When the motor is attached to the microtubule, it moves along the microtubule which was modelled as a linear filament oriented along the $x$-axes i.e. $y=0$. Assuming that the microtubule plus end was located at $x=y=0$, the effective motor placed at the origin at $t=0$ moved in the positive direction towards the minus end of the microtubule. The effective motor was modelled as a point particle connected to the cargo by a spring with a natural length $l$ and stiffness $k$ (see the illustration of the set-up in Fig.\ \ref{FIG5} (a)). The load force acting on the motor is a restoring force of the spring $\mathbf{F}_r(t)$ when it is stretched beyond $l$. Its value is given by Hooke's law, $|\mathbf{F}_r(t)| = k (d(t) - l)$, 
where $d(t)$ is the distance between the effective motor and the cargo. The force acting on the cargo is $\mathbf{F} = - \mathbf{F}_r(t)$. No force is exerted when the tether is compressed or the motor is disengaged with the filament. 
In what follows we will denote $F = |\mathbf{F}| = |\mathbf{F}_r(t)|$. The steps of $5$ nm, $8$ nm and $18$ nm were also observed experimentally (Fig B in \nameref{S1_File}). However, since they are relatively rare and do not influence the anomalous nature of the transport, in our simulations we considered only steps of $8$ nm. It is known that the dynein under load can make backward steps in the opposite direction to the direction of movement \cite{Gennerich}. We introduced a $25$\% probability for the effective motor to step in the opposite direction and also found no influence on anomalous transport phenomena. 

In our model the motor randomly binds to the microtubule with a constant attachment rate $\mathbb{T}_{a}$ and walks along the microtubule with the stepping rate which depends on the force \cite{Kunwar2011}:
\begin{equation}
\kappa(F)=%
\begin{cases}
v_0/d (1 - (F/F_s)^{1/2}), & F < F_s, \\
0,\quad & F>F_s%
\end{cases}
\label{kF}
\end{equation}%
where $v_0$ is the unloaded velocity of the single motor, $d$ is the stepping length of the motor ($8$ nm) and $F_s$ is the stall force. 
The constant attachment rate $\mathbb{T}_{a}$ in our case can be justified by the quasi one-dimentional nature of the motion. Unfortunately, in the experiment which we analyzed \cite{Kenwright}, the positions of the microtubules were not measured, therefore it was not possible to determine how far the vesicle travelled from the microtubule after the detachment. However, due to the dense nature of the cytoskeletal network, we assume that after the detachment, a new microtubule pointing in the desired direction is available.

The motor detaches from the microtubule with detachment rate $\mathbb{T}_{d}$. The observed velocity of the cargo differs from the unloaded velocity of the motor. The velocity of the cargo derives from the velocity of the motor and the friction coefficient of the cytoplasm. We chose the unloaded dynein velocity such that the cargo velocity is similar to the experimentally measured cargo velocities. For an unloaded motor velocity $v_0=4$ $\mu$m/s, we find the cargo velocity in the range of $0$ to $2$ $\mu$m/s. The typical velocity of the cargo in this case is $0.8-1.1$ $\mu$m/s. For $v_0=2$ $\mu$m/s the typical velocity of the cargo is $0.4-0.8$ $\mu$m/s. Making specific assumptions about attachment and detachment 
rates $\mathbb{T}_{a}$ and $\mathbb{T}_{d}$, we defined two models: the Markovian rate model and the non-Markovian rate model.

{\it Markovian rate model.} We considered the Markovian rate model with the attachment rate $\mathbb{T}_{a}=1$/s. The detachment rate depends on the force acting on the motor and follows an exponential dependence due to Arrhenius kinetics \cite{Kunwar2011}
\begin{equation}
\label{Td}
\mathbb{T}_{d} = \epsilon \exp(F/F_d),
\end{equation}
where $\epsilon$ is the load-free detachment rate and $F_d$ is the detachment force. 
We also considered a variant of the Markovian rate model when the detachment 
rate is constant and does not depend on the force. 
Results obtained for cargo transport in both models are similar and therefore are not reported. 

{\it Non-Markovian rate model}. For the non-Markovian rate model we take the attachment rate in the range from $\mathbb{T}_{a}=1$ s$^{-1}$  to $\mathbb{T}_{a}=100$ s$^{-1}$. The novelty of our approach is that we introduce the detachment 
rate Eq.\ (\ref{Td1}) for a effective motor defined  by inverse functions of the time interval $\tau$ between 
two detachment/attachment events. We also considered a modification of this model with a non-Markovian attachment rate 
defined similar to the Eq.\ (\ref{Td1}) (but without the force dependence). Results for this model are similar to the 
non-Markovian rate model with a constant attachment rate and are therefore not shown. 
  
Having specified the stepping behaviour, attachment and detachment kinetics of the effective motor and the force 
acting on the cargo, we described the motion of the cargo using the overdamped Langevin equation \cite{Chechkin}:
\begin{equation}
\beta \frac{d \mathbf{r}(t) }{d t} =  \mathbf{F} + \sqrt{D^{*}} \boldsymbol{\xi}_H(t),
\label{Langevin}
\end{equation} 
where $\mathbf{r}=\{x,y\}$ is the position of the cargo center of mass, $\mathbf{F}$ is the force exerted by the effective motor and $D^{*}$ is the intensity of the fGn. The stepping rate Eq. (\ref{kF}) defines how frequently the motor make $8$ nm steps. 
$\beta$ is the friction constant for dragging the cargo through its viscoelastic environment and $\boldsymbol{\xi}_H=\{ \xi_H^x,\xi_H^y \}$ is the fractional Gaussian noise (fGn). FGn is independent of the attachment and detachment kinetics of the effective motor. It is a zero-mean stationary Gaussian process and the auto-correlation function decays in the long time limit as: 
\begin{equation}
\left< \mathbf{\xi}_H(0) \mathbf{\xi}_H(t) \right> \sim 2H (2H - 1) t^{2H-2},
\end{equation}
where $H$ is the Hurst exponent, $0<H<1$. The slow decay of the auto-correlation function leads to anomalous dispersion proportional to $t^{2H}$, which for $0<H<1/2$ implies sub-diffusive behaviour. 
We note that the Langevin equation with external fGn Eq.\ (\ref{Langevin}) does not fulfill the fluctuation-dissipation relation. Therefore, there is no relation between $\beta$ and $D^{*}$. The reason why we considered the external fGn is that the medium inside living cells is highly non-equilibrium. It is well known that microrheology does not work universally inside living cells where passive viscoelastic behaviour is superposed on active phenomena due to motor proteins, see \cite{Waigh} section 4.7. The cytoskeleton is also a complex non-equilibrium network driven by cross-linking proteins and motors \cite{Mizuno}. One could not expect the FDR to be universally fulfilled in such an active environment. A different approach to sub-diffusion which is based on a fractional Langevin equation (FLE) that fulfills the FDR is pursued in Refs. \cite{Bouzat,Klein2014,GKM1,GKM2,Nam}. 

To show that our results are robust, we implemented the FLE which fulfils the FDR into our non-Markovian rate model. Results are shown in Fig.\ I in \nameref{S1_File}. In particular, for single trajectories of cargo-motor complex we plotted the distance travelled as a function of time and the time averaged variances. The results are similar to those obtained with the external fGn: the short time behaviour is sub-diffusive while on the longer time scale the behaviour is super-diffusive in TAVAR due to the non-Markovian detachment rate. Therefore we showed that the super-diffusive behaviour in the non-Markovian rate model is not affected by the nature of sub-diffusive motion. Based on these indications, we expect our other results to be robust.

In addition to fGn we also considered fluctuations which are due to errors in the measurement of the position of the cargo. These fluctuations were modelled as Gaussian noise added to the recorded coordinate of the cargo 
and lead to saturation of the TAVAR, the TAMSD  and the MFPT at small time intervals (Figs.\ \ref{FIG2}, \ref{FIG3}).

{\it Parameters of the models.} First we list the values of all parameters used in the models and discuss their choice. 
Following Ref.\ \cite{Kunwar2008}, we chose $l=100$ nm and $k=0.32$ pN/nm for the elastic tether on the cargo. The motor 
was moving via discrete increments of $8$ nm equal to the tubulin spacing along a microtubule with an unloaded velocity $v_0=4$ $\mu$m/s. The load-free detachment rate was $\epsilon=0.25$/s, 
the detachment force $F_d=3$ pN and the stall force $F_s=2.5$ pN. 
To generate the sub-diffusive behaviour in the Markovian rate and the non-Markovian rate models we simulated the fractional Gaussian noise using the exact Hosking method \cite{Hosking}. The Hurst exponent  $H=0.35$ was chosen  to match the observed sub-diffusive exponent in the experimental trajectories. The frictional coefficient was $\beta=0.72$ pN s$/\mu$m and the intensity of the fractional Gaussian noise was chosen for a single trajectory 
from the range $D^{*}=0.001-0.005$ $\mu$m$^2/$s$^{2 H}$.

\section*{Conclusion}

We have studied high quality experimental data of long range transport of individual organelles inside a living cell which is driven by molecular motors. Analysis of experimental trajectories 
revealed the anomalous nature of transport in the form of a transition from sub-diffusive behavior on short time scales to super-diffusive behavior as previously observed in Ref.\ \cite{Kenwright}. In this paper we propose a mathematical model that describes such a transition. The key ingredient of our model is the inclusion of non-Markovian kinetics: the detachment rate for an effective motor is not constant as it is commonly used for modelling, but inversely proportional to the time interval since
the last attachment event. Biologically the non-Markovian detachment rate can originate from the complex interactions between dynein motors and dynactin complexes, from interactions between separate dyneins, mixed population of dyneins
with diverse properties as speed, detachment rate or from non-Markovian internal dynamics of the dynein proteins. 
Non-Markovian internal kinetics have been discovered in a range of enzymes based on single molecule experiments \cite{Min,Chen,Xie}. We have shown that with the assumption of non-Markovian kinetics we can not only reproduce 
the experimental findings, but also there are biological implications for anomalous motor kinetics. Particularly, we have demonstrated that non-Markovian detachments increase the average distance the cargo travels in crowded environments when the 
microtubule is blocked by another filament. We also suggest that sub-diffusion could be biologically beneficial by increasing the probability of a cargo transiting from one microtubule to another, therefore promoting active motion along the microtubules. 

There is a recurrent debate as to whether the observed super-diffusive behaviour is asymptotic. From the experimental point of view it is difficult to define, since every experiment is of a finite duration, all the cells are finite sized and the microtubules have finite length. All this could restrict the longest flight duration, truncate the power law and lead to super-diffusion that appears to be transient. However, depending on the size of the cell, this transient behaviour could be very long and even last for the whole duration of the experiment. Also microtubules are grouped in bundles and the finite length of individual microtubule may play no role. In this case one would consider a transient behavior which is of the duration of the whole experiment as the asymptotic result. We did not observe any transition from super-diffusion to normal or sub-diffusion in experimental trajectories which we analysed in this work.

Finally, an interesting question arises. Dyneins have been shown to have a catch-bond mechanism when the detachment rate decreases with the increased load \cite{Kunwar2011,Rai,Mallik}. Although, in our model the non-Markovian detachment rate is a decreasing function of the time the cargo travelled along the microtubule and the force experienced by a single dynein decreases when multiple motors are engaged, it would be interesting to investigate whether the catch-bond mechanism for dyneins could lead to similar behaviour as the non-Markovian detachment rate.

\section*{Supporting information}


\paragraph*{S1 File.}
\label{S1_File}
{\bf S1 File contains 7 figures.} {\bf Fig A}, A single motor stepping in a part of experimental trajectory. 
The distance $L$ travelled by a cargo is shown as a function of time. {\bf Fig B}, The distribution of cargo 
position increments $d\textbf{r} = \textbf{r}(t +\Delta)-\textbf{r}(t)$ in experimental trajectories calculated at time
interval $\Delta=0.1$ s. The probability density function (PDF) is shown as a function of the displacement ($d\textbf{r}$).
{\bf Fig C}, Time averaged mean displacement (TAMEAN) for experimental trajectories of the lipid droplets as a function of
time interval D together with linear trend D (dashed line). {\bf Fig D}, (a) Time averaged mean squared displacement (TAMSD) for experimental trajectories of the lipid droplets as a
function of time interval D, (b) TAMSD for trajectories generated by the non-Markovian rate model and (c) by the Markovian
rate model. Power-law scaling trends are shown with $\sim \Delta^{0.7}$ (dashed-dotted line) and $\Delta^2$ (dashed line). 
{\bf Fig E}, The average velocity of a cargo manually estimated from the experimental trajectory. Parts of the trajectory with
different average velocity are marked with capital letters. The figure legend provides with the values of the average velocities
(v) corresponding to different parts of trajectory (units $\mu$m/s). {\bf Fig F}, The distribution of average cargo velocities in the experimental trajectory shown in Fig E. 
Peaks in the distribution correspond to different velocities of the cargo at parts of the trajectory indicated with capital letters next to the arrows. 
{\bf Fig G}, Distribution of average cargo velocities in two experimental trajectories (a) and (b). Peaks in the distributions
correspond to different cargo velocities. {\bf Fig H}, Distribution of average cargo velocities in trajectories generated by the non-Markovian rate model (a) and (b) and the Markovian rate model (c) and (d). 
 {\bf Fig I}, (a) Distance $L(t) = \sqrt{x^2(t)+y^2(t)}$ travelled by cargoes as a function of time for the non-Markovian rate model with
the FGE. Parameters are given in the text. (b) Time averaged variances (TAVAR) for single trajectories as a function of time
interval $\Delta$. Trajectories were of $32$ s duration. TAVAR for several longer trajectories of $72$ s duration are also shown.

\section*{Acknowledgments}
This work was supported by EPSRC Grants No. EP/J019526/1.

\nolinenumbers

%
%
%

\end{document}